# Coexistence of coupling-induced transparency and absorption of transmission signals in magnon-mediated photon-photon coupling

Biswanath Bhoi, Bosung Kim, Hae-Chan Jeon and Sang-Koog Kim[a]

*National Creative Research Initiative Center for Spin Dynamics and Spin-Wave Devices, Nanospinics Laboratory, Research Institute of Advanced Materials, Department of Materials Science and Engineering, Seoul National University, Seoul 151-744, Republic of Korea*

Coexistence of coupling-induced transparency (CIT) and absorption (CIA) of signals in magnon-mediated photon-photon coupling was experimentally determined in a planar hybrid structure consisting of a yttrium iron garnet (YIG) film and three concentric inverted-split-ring resonators (ISRRs). The experimental observation of simultaneous CIT and CIA phenomena was ascribed to magnon-mediated photon-photon coupling between the individually decoupled ISRRs. In order to capture the generic physics of the observed interactions, we constructed an appropriate analytical model based on the balance between the coherent and dissipative multiple-paths interactions, which model precisely reproduced both the CIT and CIA experimentally observed from a single hybrid system. This work, promisingly, can provide guidance for design of efficient, flexible, and well-controllable photon-magnonic devices that are highly in demand for applications to quantum technologies currently under development.

[a] Correspondence and requests for materials should be addressed to S.-K. K (sangkoog@snu.ac.kr).

The prospect of full control of electromagnetic waves has inspired intensive studies of light-matter interactions. Many novel phenomena arising from atomic coherences in light-matter interactions have potential applications to quantum information technology [1-3]. Electromagnetically induced transparency (EIT), which allows for the reduction of resonant absorption and renders transparency in a medium at/near the coupling center, is one such interesting light-matter interaction. EIT, as the backbone of future quantum memory and computation technologies, has potentially innumerable applications in optical delay generation, stoppage and light storage [4-6]. The counterpart effect of EIT is electromagnetically induced absorption (EIA), [7-8] which enables enhanced absorption in transmission spectra, thereby potentiating many applications such as fast light, photovoltaics, photodetectors, and molecular detection [8-9].

Despite broad interest in both EIT and EIA applications, devising a single system that exhibits both phenomena simultaneously has proved difficult, due to the fact that conventional coupled systems often lack independent tenability and controllability of their eigenmodes. In this regard, the hybridization of magnons (collective spin excitations) and microwave photons (electromagnetic excitations) is necessary to tailor their coupled characteristics of dispersion and damping in their common resonance region [10-14]. Furthermore, the compatibility and scalability of magnon circuits with microwave and optical lights enable magnons to be a versatile interface for a variety of quantum communication devices. For example, many studies on photon-magnon coupling (PMC) have made possible the realization of gradient memories [15], spin-current manipulation [16], nonreciprocity of microwave propagation [17], and information processing at the quantum limit [18]. In particular, PMC-mediated transparency and absorption can maximize energy conversion and/or information transfer by means of microwave-signal propagations along multiple coupling paths [19].

To explore unprecedented degrees of freedom for microwave-signal transmission and processing, in the current study we devised a robust photon-magnon hybrid system enabling coupling-induced transparency (CIT) and absorption (CIA) behaviors to coexist in a single, simple device. We also achieved substantial manipulation not only of the coupling strength but also of reliable transitions between CIA and CIT, including the normal and opposite anti-crossing dispersions. To the best of our knowledge, the present experimental demonstration shows for the first time the coexistence of CIT and CIA in a single magnon–photon coupled system, thus opening the door to magnon-mediated information transmission with preserved coherence. This work can offer not only a general approach to alternative prototyping of on-chip integrated devices but also a feasible technological solution for control and utilization of magnon-mediated photon-photon interactions that are indispensable to the development of practically simple planar-geometry quantum devices.

Figure 1(a) illustrates an experimental setup for PMC measurement with a hybrid system that consists of three concentric ISRRs and a single YIG film (see inset). The dimensions of the three different ISRRs, the common microstrip line, and YIG film are indicated in the Fig. 1(a) caption. The three concentric photon resonators on the ground plane (dark yellow) can excite microwave photon modes to couple with magnon modes excited in the YIG film (green color) placed on top of the microstrip line on the front side of the sample (see inset). We define the central point of the microstrip line as the coordinate origin, $(x, y) = (0, 0)$. Additional information on the fabrication of the concentric ISRRs and PMC measurement is provided in Refs. [12, 14].

To experimentally measure the transmission coefficient $S_{21}$, oscillating currents of microwave frequency ($f_{AC} = \omega/2\pi$) from a vector network analyzer (VNA) were applied to the microstrip feeding line at different static magnetic field $H_{dc}$ strengths applied on the x-y plane in the x-direction perpendicular to the microstrip line. Figure 1(b) shows $|S_{21}|$ power on the ($H_{dc}$ - $\omega/2\pi$) plane from only the three concentric ISRRs in the absence of YIG film. Only three peaks appeared at resonance frequencies $\omega_1/2\pi$ = 4.44 GHz, $\omega_2/2\pi$ = 5.56 GHz, and $\omega_3/2\pi$ = 7.02 GHz with intrinsic damping rates $\beta_{in_1} \simeq 2.3 \times 10^{-3}$, $\beta_{in_2} \simeq 1.8 \times 10^{-3}$ and $\beta_{in_3} \simeq 2.1 \times 10^{-3}$, respectively. The resonance frequencies and transmission characteristics (bandwidth and gain) of the three pure photon modes do not vary with $H_{dc}$. Since the resonance frequencies of the three photon modes are well separated from each other, we assume that the direct photon-photon couplings in the sample are negligible (Supplementary Materials [20], see Figs. S1 and S2). On the other hand, Fig. 1(c) shows magnon excitations in the YIG film, the so-called fundamental ferromagnetic resonance (FMR) mode as expressed by Kittel's formula $f_{FMR} = (\gamma/2\pi)\sqrt{H_{dc}(H_{dc} + \mu_0 M_s)}$, with $\mu_0 M_s$ saturation magnetization and $\gamma$ gyromagnetic ratio. The dotted line in Fig. 1(c) is the fitting of Kittel's FMR formula to the experimental data with $\mu_0 M_s$ = 0.172 T and $\gamma = 1.76 \times 10^{11}$ rad/Ts. From the widths of the FMR modes measured at different $H_{dc}$ values, the intrinsic Gilbert damping constant of the YIG film was estimated to be as small as $\alpha_{in}$ = 3.2 × 10$^{-4}$.

Then, in order to examine the interaction between the YIG and the three concentric ISRRs, we also measured the $|S_{21}|$ power from the hybrid sample of the three ISRRs and the single YIG film (inset of Fig. 1(a)). The FMR mode excited in the YIG film interacts with the first (P1), second (P2) and third (P3) photon modes at the corresponding coupling centers, as shown in Fig. 2(a). The three dispersion types of the coupled modes are very distinct in shape, though each of the

physically separated ISRRs (the same dimensions), being coupled with the YIG film, shows only the opposite anti-crossing (Supplementary Materials [20], see Fig. S3). The magnified images at the corresponding coupling centers (Fig. 2(b)) represent the distinct types, of normal anti-crossing, absorption, and opposite anti-crossing, at 7.02, 5.56, and 4.4 GHz, respectively. In detail, the normal anti-crossing shows a level repulsion between the higher- and lower-frequency branches (Fig. 2(b): right); the opposite anti-crossing corresponds to a level attraction between the higher- and lower-frequency branches at the crossing center (see Fig. 2b, left). These two types of anti-crossing represent microwave transmissions with minimal absorption at the corresponding coupling centers, like CIT. Unlike these two anti-crossing types, the dispersion shown in the middle of Fig. 2(b) is strikingly different: the two hybridized modes were attracted and converged to a single frequency level with relatively strong microwave absorption at the coupling center, the so-called CIA. Quite recently, such similar absorption phenomena were predicted for synthetic layered magnets [21], but have yet to be found experimentally in photon-magnon hybrid samples.

In order to elucidate the coexistence of CIT and CIA in a single hybrid device, we proposed magnon-mediated coupling between the uncoupled photon modes, in which coupling the individual photon modes are coupled directly to a single magnon mode, and then their hybridized modes interact with each other, as schematically shown in Fig. 3. We used an electrodynamic approach to create an appropriate model of magnon-mediated photon-photon coupling, employing Faraday's and Ampère's laws for coherent interaction and Lenz's law for dissipative interaction, as well as the Landau-Lifshitz-Gilbert equation for magnon dynamics, as described in Refs. [14,19]. For the direct interaction (indicated by the solid double-arrow gray lines in Fig. 3) of the magnon with each of the three photon modes, n=1, 2, and 3, the eigenvalue equations can be expressed in matrix form as (Supplementary Materials [20], see Sec. S4)

$$\begin{bmatrix} \omega - \widetilde{\omega}_r & \omega_m K_1^2 & \omega_m K_2^2 & \omega_m K_3^2 \\ \omega_1 & \omega - \widetilde{\omega}_1 & 0 & 0 \\ \omega_2 & 0 & \omega - \widetilde{\omega}_2 & 0 \\ \omega_3 & 0 & 0 & \omega - \widetilde{\omega}_3 \end{bmatrix} \begin{bmatrix} m^+ \\ J_1^+ \\ J_2^+ \\ J_3^+ \end{bmatrix} = 0 \qquad (1)$$

where $\widetilde{\omega}_r = \omega_r - i\alpha_{in}\omega_r$ and $\widetilde{\omega}_n = \omega_n - i\beta_{in_n}\omega_n$ are the complex frequencies of the magnon and individual photon modes, respectively, and $\alpha_{in}$ and $\beta_{in_n}$ are the intrinsic damping parameters for the magnon and $n^{th}$ photon modes, respectively. The energy of the magnons interacting with each of the individual photon modes can be modified into damping as an additional dissipation channel. Therefore, the effective damping constant for the magnon is expressed as $\alpha_{eff} = \alpha_{in} + \alpha_{cp}$ with an additional constant of coupling-induced damping $\alpha_{cp}$. $K_n$ represents the coupling constant between the magnon and the $n^{th}$ photon mode; $m^+$ and $J_n^+$ are the dynamic magnetizations in the YIG and the current density in the $n^{th}$ ISRR resonator, respectively (Supplementary Materials [20], see Sec. S4). For the case of normal anti-crossing, the coupling constant has a real value of $K$ (i.e. coherent coupling), whereas for opposite anti-crossing, $K$ has an imaginary value (i.e. dissipative coupling). It is known [14] that the normal and opposite anti-crossing is determined by the relative strength and phase of the oscillating magnetic fields generated from both the microstrip line and the ISRRs.

For the above-noted direct interaction of the magnon with each of the three photon modes (n=1, 2, and 3), the eigenfrequencies ($\omega_{n\pm}$) of the higher (+) and lower (-) branches of the two hybridized modes can thus be written analytically by solving Eq. (1), as

$$\omega_{n\pm} = \left[\frac{(\widetilde{\omega}_r + \widetilde{\omega}_n)}{2} \pm \frac{1}{2}\sqrt{(\widetilde{\omega}_r - \widetilde{\omega}_n)^2 + 2\omega_m\omega_n K_n^2}\right] \qquad (2)$$

Among the hybridized modes expressed as $\omega_{n\pm}$, the higher branch ($\omega_{1+}$) of the P1 photon-magnon (P1-M) coupling is coupled to the lower branch ($\omega_{2-}$) of the P2-M coupling, thus resulting in indirect coupling between the P1 and P2 photon modes via their direct coupling with the magnon mode. Similarly, the $\omega_{2+}$ mode is coupled to the $\omega_{3-}$ mode, thus leading to indirect coupling between the P2 and P3 modes, and finally, the $\omega_{3-}$ mode is coupled to the $\omega_{1+}$ mode, resulting in indirect coupling between the P3 and P1 modes. Therefore, the effective coupling constant $K_n$ for the magnon-mediated photon-photon coupling is modified as $K_n^2 = k_{MP_n}^2 + k_{PP}^2$, where $k_{MP_n}$ is the direct PMC constant and $k_{PP}$ is the magnon-mediated, indirect photon-photon coupling constant as given by $k_{pp}^2 = k_{P_1P_2}^2 + k_{P_1P_3}^2 + k_{P_2P_3}^2$. The value of $k_{PP}$ primarily depends on the relative amplitude ($\sigma$) and phase difference ($\psi$) between the individual pure photon modes. However, since the modes $\omega_{1-}$ and $\omega_{2-}$ move away from the upper hybridized mode $\omega_{3+}$, there will not be any contribution to $k_{PP}$. The numerical values $k_{MP_n}$ were experimentally estimated from each hybrid sample composed of YIG and a single nth ISRR (not the three concentric ISRRs) (Supplementary Materials [20], see Fig. S3 and Sec. S3).

Fitting of the real part of Eq. (2) to the experimentally observed dispersions shown in each of the Fig. 2 (b) plots (opposite anti-crossing, absorption, normal anti-crossing, respectively) led to the higher- and lower-frequency branches of the corresponding hybrid modes (solid black lines), which are in good agreement with the experimental data. From the fitting results, the net coupling constant for the opposite anti-crossing at 4.48 GHz was found to be $K_1 = 0.008i$ with $\alpha_{cp_1} \sim 0$.

Since $K_1$ is equal to the value estimated for the direct P1-M coupling ($k_{MP_1} = 0.008i$) from the hybrid sample of the single 1st ISRR and YIG, the $k_{PP}$ was negligible for the case of the opposite anti-crossing at 4.48 GHz. The imaginary value of $K_1$ indicates that the P1-M coupling is dominated by a dissipative coupling due to a strong microwave magnetic field generated from the 1st ISRR [12]. The fitting of Eq. (2) to the P2-M coupling dispersion (CIA) yielded $K_2 = 0.004i$ with $\alpha_{cp_2} = 0.01$. The estimated value of $\alpha_{cp_2}$ was much greater than the intrinsic damping ($\alpha_{in} = 3.2 \times 10^{-4}$) of the magnon mode. This means that the magnon-mediated interaction provides a two-tone decay, where an initial rapid decay by the extrinsic (coupling-induced) damping to the lowest energy state is followed by a slower collective decay due to the intrinsic damping. By adopting $k_{MP_2} = 0.008i$ estimated experimentally from the hybrid sample composed of a single 2nd ISRR and YIG (Supplementary Materials [20], see Sec. S3), the magnon-mediated photon-photon coupling constant was found to be $k_{PP} = 0.005$ for the CIA dispersion at 5.38 GHz. Lastly, the fitting to the normal anti-crossing dispersion at 6.99 GHz led to $K_3 = 0.01$ and $\alpha_{cp_3} = 2 \times 10^{-4}$. Compared with the $\alpha_{cp_2}$ (due to P2-M coupling), $\alpha_{cp_3}$ is too small to be negligible. From $k_{MP_3} = 0.008i$ estimated from the single 3rd ISRR/YIG hybrid sample (Supplementary Materials [20], see Sec. S3), the magnon-mediated photon-photon coupling constant for the normal anti-crossing was found to be $k_{PP} = 0.013$. Accordingly, the observation of both CIT (normal and opposite anti-crossing) and CIA can be ascribed to the balance between the coupling multiple paths of direct photon-magnon and magnon-mediated indirect photon-photon interactions. The input signal sustained by the indirect photon-photon coupling through the magnon-mediated interactions encounters a partial destructive interference with the input signal by the direct photon-magnon coupling In other words, the competition between coherent $k_{PP}$ and dissipative $k_{MP}$ determines the types of anti-crossing at the different coupling centers in the three-concentric-ISRR/YIG hybrid sample. Thus, we found the opposite anti-crossing dispersion for P1-M coupling ($|k_{MP}| > |k_{PP}|$) and the normal dispersion for P3-M coupling ($|k_{MP}| < |k_{PP}|$). On the other hand, for a system with dissipative coupling equal to or slightly higher than coherent coupling ($|k_{MP}| \approx |k_{PP}|$), along with the mixed relaxation channels of the magnon mode, microwave power can be absorbed at the corresponding coupling center, and thus CIA can result, as shown for P2-M coupling.

In order to examine the occurrence of different types of dispersion spectra due to magnon-mediated photon-photon coupling, as discussed above for the concentric ISRR/YIG hybrid system, we numerically calculated the $|S_{21}|$ power on the $\omega/2\pi$ - $H_{dc}$ plane for the three different coupling regions using $|S_{21}|_n = \Gamma_n \omega^2 (\omega - \widetilde{\omega}_r)/\left[(\omega - \widetilde{\omega}_r)(\omega - \widetilde{\omega}_n) - \frac{1}{2}\omega_m \omega_n (k_{MP}^2 + k_{PP}^2)_n\right]$ as shown in Fig. 4(a) – (c). In this numerical calculation, we used the values estimated from the experimental results and $\Gamma_n \approx 2\beta_n$ represents the ISRR/YIG hybrid/cable impedance mismatch. The calculation results agree well with the three different experimentally observed dispersions, i.e., normal (7.02 GHz) and opposite (4.4 GHz) anticrossing, and CIA (5.58 GHz) dispersion. To gain further insight into and understanding of the similarity and/or difference between the observed CIT and CIA behaviors, we employed Eq. (2) to numerically calculate the complex eigenvalues for two split branches and the three different types of dispersions, $E_\pm = \omega_\pm - \Delta\omega_\pm$ with $\omega_\pm$ the angular frequency (blue solid line) and $\Delta\omega_\pm$ the linewidth (red dashed line) of the upper and lower branches of each hybridized mode, as shown in Figs. 4(d) – 4(f). For the case of coherent coupling (normal anti-crossing), the hybridized mode's frequencies are repelled with the cross-over of their linewidths (Fig. 4(f)), while for the dissipative coupling (opposite anti-crossing), they show the opposite trend (Fig. 4(d)); nonetheless, both cases exhibit CIT at each coupling center. On the other hand, Fig. 4(e) represents CIA behavior, wherein the frequency dispersion shows cross-over behavior, and the bifurcation points are almost invisible (Fig. 4(e)). This type of dispersion can also be observed in a very weakly coupled photon-magnon hybrid, making it difficult to distinguish between CIA and no coupling (crossing). However, even in cases where split frequencies in CIA cross each other, there exists strong coupling between the ISRR and the magnon mode at/near the coupling center, which leads to strong microwave absorption (microwave transmission blocking). Furthermore, the linewidths of the split branches are still repulsive within a narrow field range (see Fig. 4(e), center), which manifestation is distinct from the crossing behavior of weakly coupled hybridized modes.

Further to the above-noted experimental and analytical findings, an important consequence of the simultaneous exhibition of both CIT and CIA is the subsuming of $\alpha_{cp}$ due to a cooperative effect of direct photon-magnon and magnon-mediated photon-photon interactions that depend on the local electromagnetic environment of the photon modes. Eventually, the relative amplitude ($\sigma$) and phase difference ($\psi$) for the magnon-mediated photon-photon coupled modes affect $\alpha_{cp}$, which leads the system to absorb microwave power at the coupling center. To better understand the role of magnon-mediated parameters ($\sigma$ and $\psi$) with $k$ and total damping $|\beta - \alpha_{\text{eff}}|$ in CIT/CIA, we derived a generalized analytical form for the frequency gap ($\Delta$) between the hybridized modes at the coupling centers associated with magnon-mediated photon-photon coupling, as $\Delta = \frac{\omega_p}{2\pi}\sqrt{-2k^2(1 - \sigma cos\psi) - (\beta - \alpha_{\text{eff}})^2}$ (Supplementary Materials [20], see Sec. S6). Figure 5(a) shows a phase diagram of $\Delta$ for the occurrence of CIA/CIT on the plane of $\psi$ and $|\beta - \alpha_{\text{eff}}|$ for constant values of both $k = 0.02$ and $\sigma = 2$. The two different types of anti-crossing for CIT are distinguished by the condition of $\Delta = 0$ i.e. $\psi = cos^{-1}\left[\frac{1}{\sigma}\left(1 + \frac{(\beta-\alpha_{\text{eff}})^2}{2k^2}\right)\right]$, as indicated by the black dotted line in Fig. 5(a). The $\Delta = 0$ criterion represents CIA occurrence. Based on the phase diagram, we can summarize the CIA and CIT features in our coupled system as follows: (as noted by the colors in the different regions) $\Delta$ in the CIT regions gradually decreases as it moves closer to the CIA line. As $\psi$ increases in the range of $|\beta - \alpha_{\text{eff}}| > 0.03$, the anti-crossing in CIT becomes the opposite type with a weak value of $\Delta$, whereas as $\psi$ increases in the range of $|\beta - \alpha_{\text{eff}}| < 0.03$ below the marked boundary curve, the anti-crossing in CIT becomes a normal one with increasing $\Delta$. When the damping rates of $\alpha$ and $\beta$ are highly mismatched, the signature of CIA becomes less clear. Further, the CIA line or the boundary between the two CIT regions shifts toward higher $\psi$ values with the increase in $\sigma$ for a constant value of $k = 0.02$, as shown in Fig. 5(b).

Similarly, the CIA line shifts toward higher $\psi$ values with the increase in $k$ for a constant value of $\sigma = 2$, as shown in Fig. 5 (c). It is also evident that the dissipative coupling can be converted to a coherent one or vice versa by controlling the strength and phase of the photon modes that indirectly interact via magnons. Therefore, the creation of dispersion types during coupling is more dependent on the local properties of the photon (i.e. strength and phase) and magnon modes (i.e. coupling-induced damping) than the global properties of the intrinsic damping and the dimensions of the coupled resonators.

In conclusion, by designing a unique planar hybrid device of three concentric ISRRs and a single YIG film, we demonstrate the coexistence of CIT with CIA in magnon-mediated photon-photon coupling. The cooperation of the coupling-induced damping of magnons as well as the interference between the multiple pathways of magnon-mediated photon-photon interactions and direct photon-magnon interaction governs the key physics for the new scheme of CIA and CIT expression in a photon-magnon hybrid system. The proposed analytical model based on the cooperative effect of direct photon-magnon and magnon-mediated photon-photon interactions supports our experimental results, which an help to widen applications of PMC for multiple signal processing. Furthermore, in spintronics, the tenability of CIT/CIA from a photon-magnon hybrid system can provide a means of transmitting and manipulating spin excitations coherently at long distances, particularly with photon excitations outperforming the currently reported micrometer propagation using pure spin currents or spin waves. Thus, this demonstration of multifunctional characteristics of PMC in a single planar device can be a crucial stepping stone to the development of more complex, controllable and sensitive quantum information processing devices.

**ACKNOWLEDGMENTS**

This research was supported by the Basic Science Research Program through the National Research Foundation of Korea (NRF) funded by the Ministry of Science, ICT, and Future Planning (No. NRF-2021R1A2C2013543). The Institute of Engineering Research at Seoul National University provided additional research facilities for this work.

# Figure captions

FIG. 1. (a) Schematic drawing of measurement setup of PMC from hybrid system composed of YIG film (green) and three concentric ISRRs (shown in inset). Both end ports of the microstrip feeding line are connected to a VNA, and static magnetic fields ($H$) are applied along the x-axis using an electromagnet. The dimensions of the concentric ISRRs are as follows: $a$ = 6 mm, b = 5 mm, and c = 4 mm with equal values of $s$ = 0.3 mm and $g$ = 0.25 mm. The size of the YIG film is 3.7 mm × 3.7 mm × 25 μm. The width and thickness of the microstrip line are $w$ = 1.58 mm and $t_c$ = 35 μm, respectively. The thickness of the dielectric substrate is $t_d$ = 0.74 mm. (b) and (c) show |$S_{21}$| power experimentally measured as functions of microwave AC frequency ($f_{AC} = \omega/2\pi$) and static field strength ($H_{dc}$) separately from only (b) the three ISRRs and (c) the YIG film. The dotted line in (c) is a result of fitting to the experimental data using Kittel's FMR formula.

FIG. 2. (a) |$S_{21}$| power contour plot measured as both functions of $\omega/2\pi$ and $H_{dc}$ from three concentric ISRRs/YIG hybrid sample, as noted in Fig. 1(a). 'P1', 'P2', and 'P3' represent the three different photon modes at the corresponding resonance frequencies. (b) Magnified contour plots of |$S_{21}$| power for three coupling dispersions denoted as ' P1', 'P2', and 'P3' photon modes being coupled with magnon mode excited in YIG film. The black solid lines correspond to the calculated dispersions according to Eq. (2) shown in the text.

FIG. 3. Schematic drawing illustrating mechanism for magnon-mediated photon-photon coupling between initially decoupled photon modes ($\omega_n$). The solid double-arrow gray lines represent the coupling of the FMR magnon mode ($\omega_r$) with each of the individual three-photon modes, $\omega_1$, $\omega_2$ and $\omega_3$, which direct couplings result in two split modes, the higher-frequency $\omega_{n+}$ and lower-

frequency $\omega_{n-}$ hybrid modes for each photon mode $\omega_n$. Then, the higher modes $\omega_{n+}$ can couple with the lower modes $\omega_{(n+1)-}$ of the next photon modes, as denoted by the dotted double-arrow lines. This coupling mechanism represents magnon-mediated, indirect photon-photon couplings.

FIG. 4. Numerically calculated $|S_{21}|$ power on the $\omega/2\pi$ -$H_{dc}$ plane for three different coupling regions of the concentric three ISRRs/YIG hybrid system according to magnon-mediated photon-photon coupling: (a) opposite anti-crossing in CIT (b) absorption, CIA and (c) normal anti-crossing in CIT. (d), (e), and (f) show the three different analytically calculated dispersions as represented by frequency ($\omega$) (solid blue lines) and linewidth ($\Delta\omega$) (dashed red lines) versus static magnetic field ($H_{dc}$).

FIG. 5. (a) Analytically calculated phase diagram of coupling dispersion types on plane of $\psi$ - $|\beta - \alpha_{eff}|$ for case of $k = 0.02$ and $\sigma = 2$. The color indicates the absolute value of net coupling strength $\Delta$ between the two split branches of the corresponding hybrid modes. $\Delta = 0$ indicates a boundary (CIA) that distinguishes the normal and opposite anti-crossing in the CIT. The boundary (CIA line) on the $\psi$ - $|\beta - \alpha_{eff}|$ plane varies with (b) $\sigma$ for $k = 0.02$ (as indicated by the red curve), and (c) $k$ for $\sigma = 2$ (as indicated by the blue curve).

**Fig. 1**

Fig. 1

**Fig. 2**

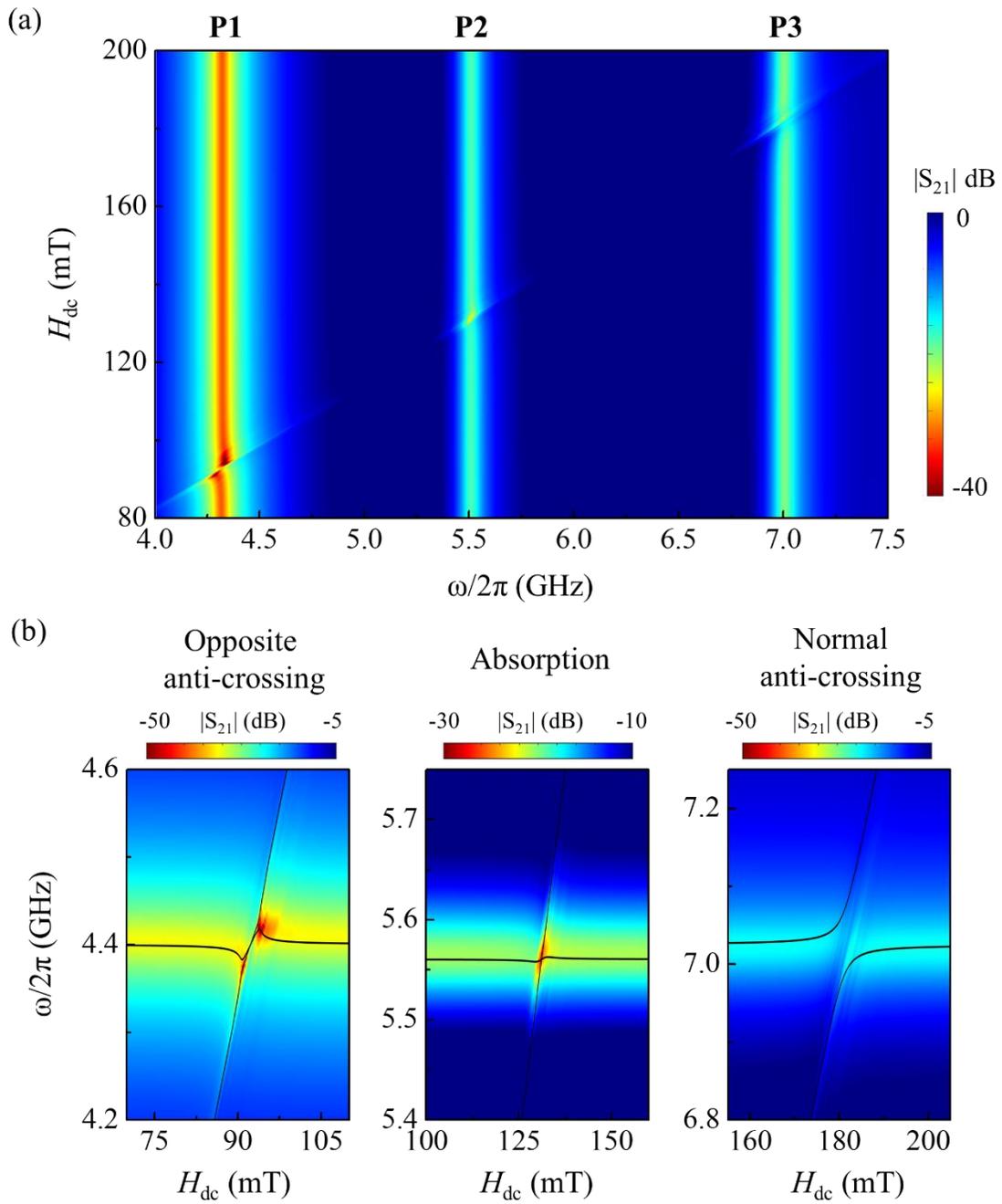

**Fig. 3**

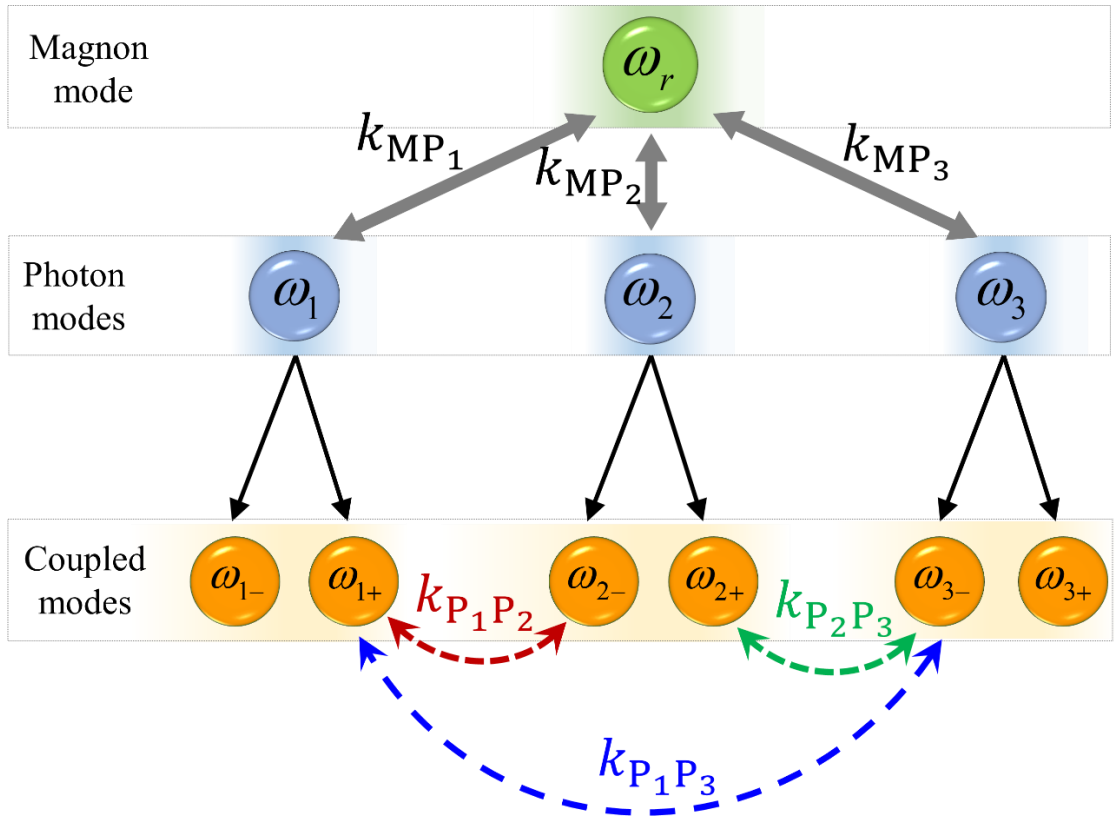

**Fig. 4**

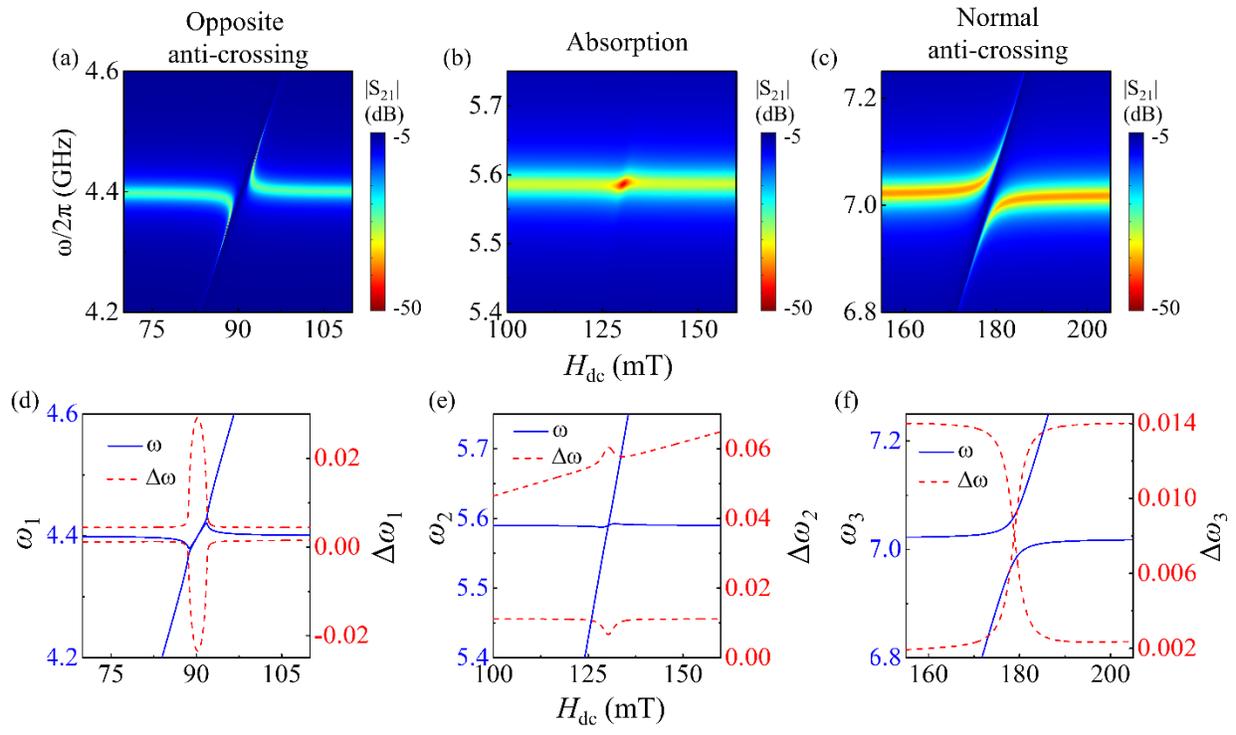

**Fig. 5**

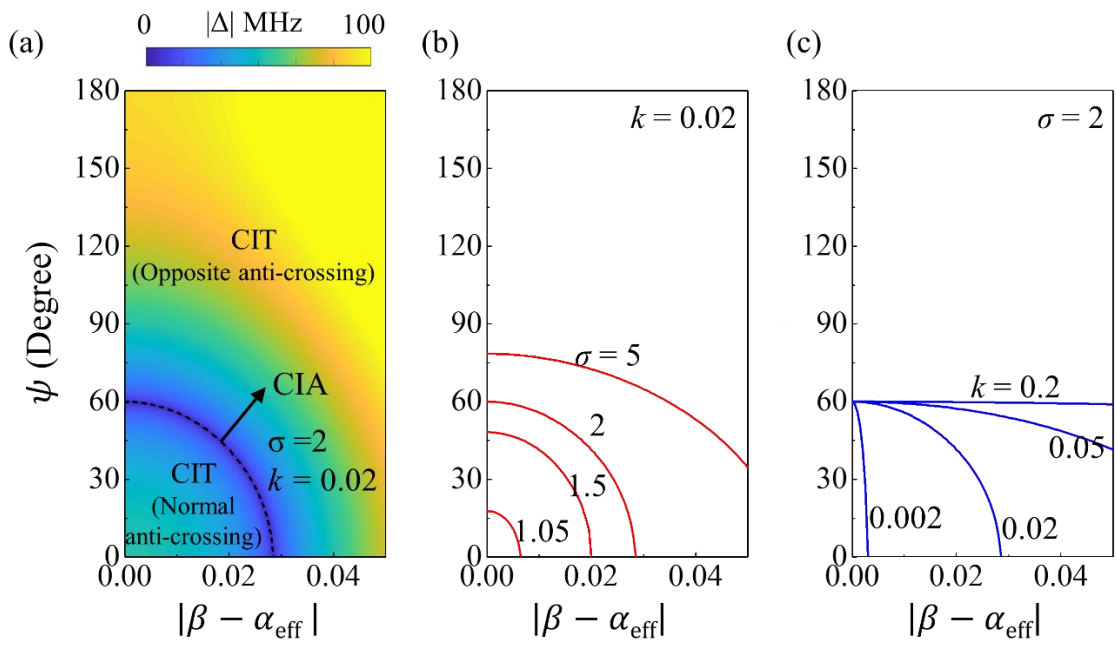

**SUPPLEMENTARY MATERIALS**

# Coexistence of coupling-induced transparency and absorption of transmission signals in magnon-mediated photon-photon coupling

Biswanath Bhoi, Bosung Kim, Hae-Chan Jeon and Sang-Koog Kim[a]

*National Creative Research Initiative Center for Spin Dynamics and Spin-Wave Devices, Nanospinics Laboratory, Research Institute of Advanced Materials, Department of Materials Science and Engineering, Seoul National University, Seoul 151-744, Republic of Korea*

[a] Correspondence and requests for materials should be addressed to S.-K. K (sangkoog@snu.ac.kr).

## S1. (a) Numerical simulations of |$S_{21}$| for each of 1st, 2nd, 3rd individual ISRRs

The numerical simulations for three concentric ISRRs (Fig. S1(a)) and each of the individual ISRR components (Fig. S1(b)) were performed using a commercial electromagnetic full-wave simulator (CST Microwave Studio). The dimensions of each of the ISRRs and the microstrip line are indicated in their respective figures. All the samples had an equal thickness of microstrip line, 35 μm, and a dielectric substrate of 0.74 mm.

Transmission spectra |$S_{21}$| as functions of AC current frequency obtained from numerical simulations for the three concentric ISRRs and for each of the individual ISRR components are shown in Figs. S1(c) and S1(d), respectively. The resonance frequencies for the three concentric ISRR peaks are 4.5, 5.37 and 6.94 GHz, very close to the resonance frequencies 4.51, 5.61, and 7.1 GHz, respectively, corresponding to each of the ISRR components. Furthermore, the resonance frequencies of the three concentric ISRRs in the sample are well separated compared with those of the ISRR components, indicating that coupling between the three photon modes is negligible.

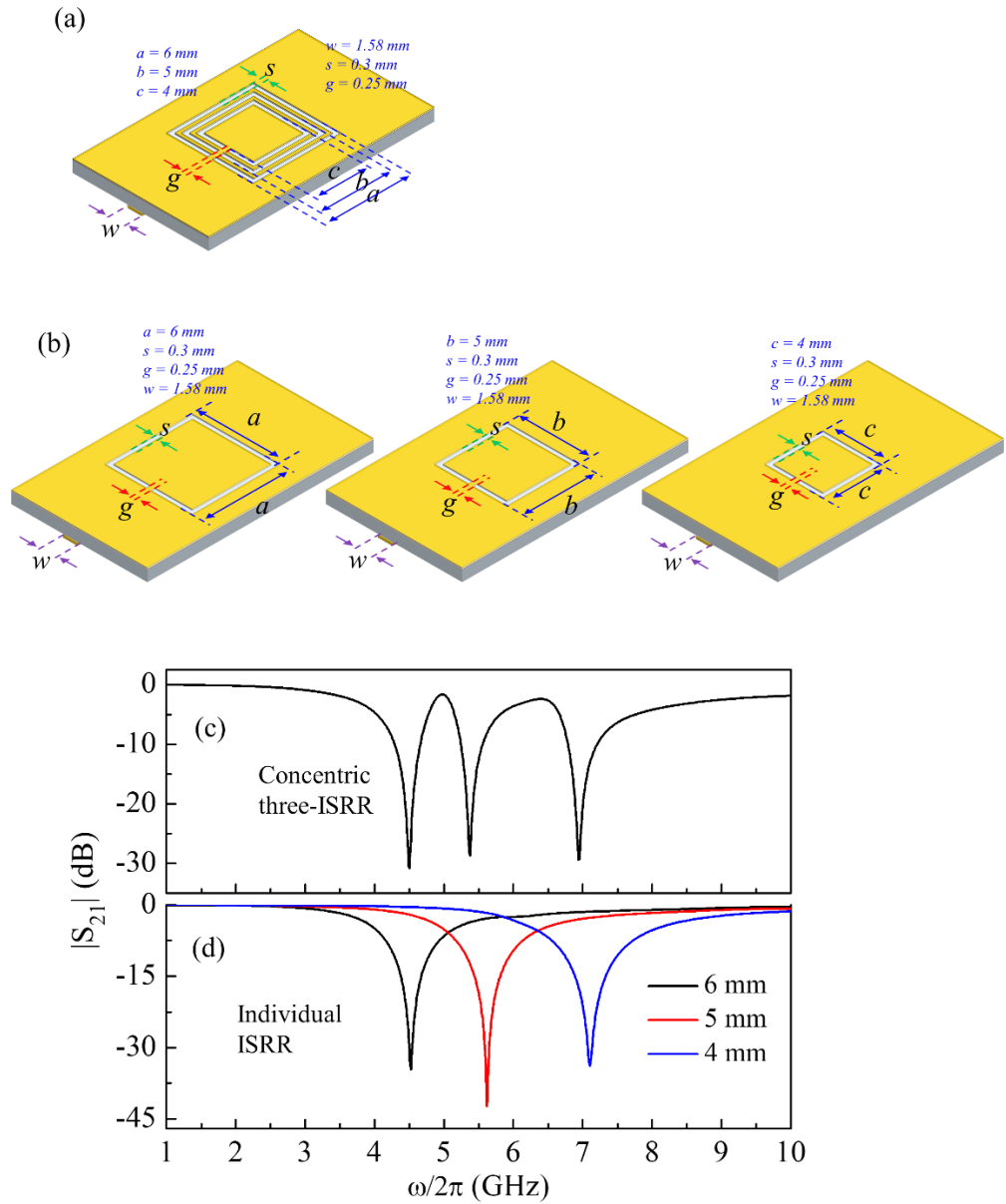

**Fig. S1:** Schematic view of ISRR structures with different geometrical parameters: (a) three concentric ISRRs and (b) individual ISRR. Transmission spectra |S$_{21}$| as function of microwave frequency for (c) three concentric ISRRs and (d) individual ISRRs.

**S1 (b). Experimentally measured magnitude and phase of |S$_{21}$| for three concentric ISRRs**

Figure S2 shows the experimentally measured magnitude (black solid line) and phase (blue solid line) of transmission spectra |S$_{21}$| as a function of ac current's frequency from the three-concentric-ISRR sample without YIG on the microstrip line. Similar to its numerical simulations results, the sample shows three resonance peaks P1, P2 and P3 at 4.44, 5.56 and 7.02 GHz respectively. The magnitude and phase corresponding to each resonance peak P1, P2 and P3 were found to be ($\sigma_1$ = -33.0 dB, $\psi_1$= 70°), ($\sigma_2$ = -20.8 dB, $\psi_2$= 170°) and ($\sigma_3$ = -22.5 dB, $\psi_3$= -320°), respectively. The phase difference between the photon modes is given as $\psi_{21}$=100°, $\psi_{31}$=250° and $\psi_{32}$=150°, respectively, where the subscript represents the mode number.

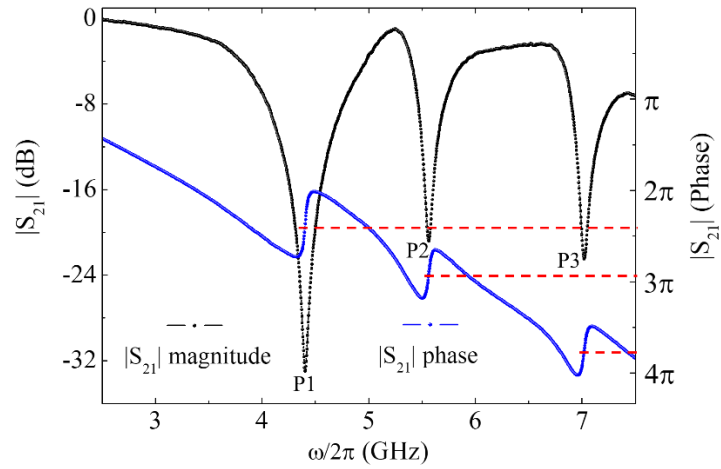

**Fig. S2:** Experimentally measured magnitude and phase of transmission spectra |S$_{21}$| as functions of ac current's frequency for three-concentric-ISRR sample without YIG film on microstrip line and in absence of bias magnetic field

## S2. Theoretical formalism for single photon-magnon coupling (PMC)

The PMC in an ISRR/YIG hybrid system is based on the combination of a microwave LCR and Landau-Lifshitz-Gilbert (LLG) equations. The coupling of the magnetization dynamics with a microwave is established through an electrodynamic approach by the combined action of Ampere's law, Faraday's law and Lenz's law.

A static magnetic field $H_{dc}$ created by an electromagnet is applied, in the x direction, to an ISRR/YIG hybrid system lying on the *x-y* plane. This hybrid system consists of three different physical systems: 1) the microstrip line to excite the magnon and photon modes as well as to probe those coupled modes; 2) the YIG, and 3) the ISRR wherein the magnon and photon modes are to be excited. When ac currents (*j*) are applied along the microstrip line placed on the *y*-axis, an electromotive force (EMF) voltage $V$ is generated in the ISRR, as expressed by $V = Z_p j$, $Z_{ISRR}$ being the ISRR's impedance, which is given, according to an equivalent LCR circuit model, as

$$Z_p = -\frac{iL}{\omega}(\omega^2 - \omega_p^2 + 2i\beta\omega\omega_p), \tag{S1}$$

where $\omega_p = 1/\sqrt{LC}$ is the angular resonance frequency of ISRR with inductance $L$ and capacitance $C$, and $\beta$ is the damping parameter of ISRR.

For the YIG only, the oscillating magnetic field created (via Ampere's circuit law) by the ac currents flowing along the microstrip line can directly stimulate magnetization excitations in the YIG film, as described by the Landau-Lifshitz-Gilbert (LLG) equation [S1-S2]

$$\frac{d\boldsymbol{m}}{dt} = -\gamma \boldsymbol{m} \times \boldsymbol{H}_{\text{eff}} + \alpha \boldsymbol{m} \times \frac{d\boldsymbol{m}}{dt}, \tag{S2}$$

where $\boldsymbol{m} = \boldsymbol{M}/M_s$ is the magnetization vector, with gyromagnetic ratio $\gamma/2\pi = 28$ GHz/T, intrinsic

Gilbert damping parameter $\alpha = 3.2 \times 10^{-4}$ and saturation magnetization $\mu_0 M_s = 0.172$ T, as derived from the FMR measurement of only the YIG film. $\boldsymbol{H}_{\text{eff}}$ is the effective magnetic field, given as $\boldsymbol{H}_{\text{eff}} = \boldsymbol{H}_{\text{dc}} + \boldsymbol{h}e^{-i\omega t}$, where $\boldsymbol{H}_{\text{dc}} = H\hat{\boldsymbol{x}}$ is the static magnetic field externally applied in the x-direction and $\boldsymbol{h}e^{-i\omega t}$ is the ac magnetic field generated from the microstrip line with amplitude $|\boldsymbol{h}|$ and angular frequency $\omega$. Using a linearized form of the magnetization direction, the magnetization variation is given as $\boldsymbol{m} \cong M_s\hat{\boldsymbol{x}} + \boldsymbol{m}_\perp e^{-i\omega t}$, where $\boldsymbol{m}_\perp e^{-i\omega t}$ is the oscillating component of the magnetization on the y-z plane. Assuming $|\boldsymbol{m}_\perp| \ll M_s$, the LLG equation can be simplified in the rotational frame to

$$(\omega - \omega_r + i\alpha\omega)m^+ + \omega_m h^+ = 0, \quad (S3)$$

where $m^+ = m_y + im_z$, $h^+ = h_y + ih_z$ and $\omega_r = \gamma\sqrt{H_{\text{dc}}(H_{\text{dc}} + \mu_0 M_s)}$ is the ferromagnetic resonance frequency of YIG film, and $\omega_m = \gamma\mu_0 M_s$.

Once the magnetizations are excited in the YIG, they can yield an additional voltage to the ISRR according to Faraday's induction law, as given by $V_y = k_F L(dm_z/dt)$ and $V_z = k_F L(dm_y/dt)$. The total induced voltage in the ISRR is thus $V_{p\leftarrow m} = V_y + iV_z = -k_F L\omega m^+$, where $k_F$ is the coupling parameter due to Faraday's induction law. This induced voltage generates an additional microwave current (Lenz's law) in the ISRR, as expressed by $V_{p\leftarrow m} = Z_p J^+$, where $J^+ = J_y + iJ_z$. Using Eq. (S1), this relation is finally written as

$$ik_F\omega^2 m^+ + \left(\omega^2 - \omega_p^2 + 2i\beta\omega\omega_p\right)J^+ = 0. \quad (S4)$$

Due to the effect of Lenz's law, the voltage $V_{p\leftarrow m}$ induced by Faraday's law will generate an induced current in the ISRR and hence an additional magnetic field will be created around the ISRRs' split gap. Thus, the magnetizations in the YIG are influenced by two time-dependent magnetic fields: one generated by the current in the microstrip line that drives the magnetization dynamic (Ampère's law), and the other generated by the induced current in the ISRR, which impedes the magnetization dynamic (Lenz's law). Taking into account the total magnetic field that contributes to the magnetization excitation in the YIG film, the LLG equation (Eq. S3) in the rotating frame is thus rewritten as

$$(\omega - \omega_r + i\alpha\omega)m^+ - i\omega_m(k_A - k_L)J^+ = 0, \tag{S5}$$

where $J^+ = i(h^+)_p/(k_A - k_L)$ is the net microwave current in the ISRR circuit, resulting, via Ampere's and Lenz's laws, in the magnetic field $h_p$: $(h_y)_p = (k_A - k_L)J_z$ and $(h_z)_p = -(k_A - k_L)J_y$, where $k_A$ and $k_L$ are the coupling parameters due to Ampere's and Lenz's laws, respectively. To obtain the simultaneous solutions of Eqs. (S4) and (S5), the matrix form is rewritten as

$$\begin{pmatrix} \omega^2 - \omega_p^2 + 2i\beta\omega\omega_p & ik_F\omega^2 \\ -i\omega_m(k_A - k_L) & \omega - \omega_r + i\alpha\omega \end{pmatrix} \begin{pmatrix} J^+ \\ m^+ \end{pmatrix} = \begin{pmatrix} 0 \\ 0 \end{pmatrix} \tag{S6a}$$

$$\Omega \begin{pmatrix} J^+ \\ m^+ \end{pmatrix} = \begin{pmatrix} 0 \\ 0 \end{pmatrix} \tag{S6b}$$

The determinant of $\Omega$ is expressed as $(\omega - \omega_r + i\alpha\omega)(\omega^2 - \omega_p^2 + 2i\beta\omega\omega_p) - \omega^2\omega_m k_F(k_A - k_L) = 0$, $K^2 \cong k_F(k_A - k_L)$; as such, it finally describes PMC in the single ISRR/YIG hybrid system, $K$ being the net coupling constant between the magnon mode and the photon mode. For the case of $k_A > k_L$, normal anti-crossing is observed with real $K$ (i.e. coherent coupling), whereas for the case of $k_A < k_L$, opposite anti-crossing is found with imaginary $K$ (i.e. dissipative coupling). Whether the normal or opposite anti-crossing appears is determined by the relative strength and phase of the oscillating magnetic field between the ISRR and the microstrip line, both of which can excite magnons in the YIG film. The equation of motion for a single PMC can be written in a generalized form as

$$(\omega - \widetilde{\omega}_r)(\omega - \widetilde{\omega}_p) - \frac{1}{2}\omega_m\omega_p K^2 = 0, \tag{S7}$$

where $\widetilde{\omega}_r = \omega_r - i\alpha\omega_r$ and $\widetilde{\omega}_p = \omega_p - i\beta\omega_p$ are the complex frequencies of the magnon and photon modes, respectively. Eq. (S7) can be solved to obtain the complex eigenvalues of the hybridized modes:

$$E_\pm = \frac{1}{2}\left[(\widetilde{\omega}_r + \widetilde{\omega}_p) \pm \sqrt{(\widetilde{\omega}_r - \widetilde{\omega}_p)^2 + 2\omega_m\omega_p K^2}\right] \tag{S8}$$

Using input–output formalism, we can further obtain the transmission spectrum for a single photon-magnon coupled system as

$$S_{21} = \Gamma \frac{\omega^2(\omega - \widetilde{\omega}_r)}{(\omega - \widetilde{\omega}_r)(\omega - \widetilde{\omega}_p) - \frac{1}{2}\omega_m\omega_p K^2} \tag{S9}$$

where $\Gamma$ represents the ISRR/YIG hybrid/cable impedance mismatch.

## S3. Experimental measurements of three different hybrid samples composed of YIG and each of the ISRR components

We measured $|S_{21}|$ power from the individual hybrid samples composed of YIG and a single ISRR of 6, 5 or 4 mm. The FMR mode excited in the YIG film interacts with each ISRR's photon mode, resulting in opposite anti-crossing dispersions (level attraction between the split modes) for all three hybrid samples, as shown in Fig. S3. The coupling constant ($k_{\text{MP}}$) for each of the hybridized modes due to PMC is estimated by fitting the real part of Eq. (S8) to the lower- and higher-frequency branches (the black solid lines shown in Fig. S3) of the dispersion spectra. Interestingly, for all of the three-ISRR/YIG hybrid samples, the coupling constant was found to be the same, $K = k_{\text{MP}_n} = 0.008i$. Thus, in this case, the coupling constant is likely to be independent of the ISRR dimensions.

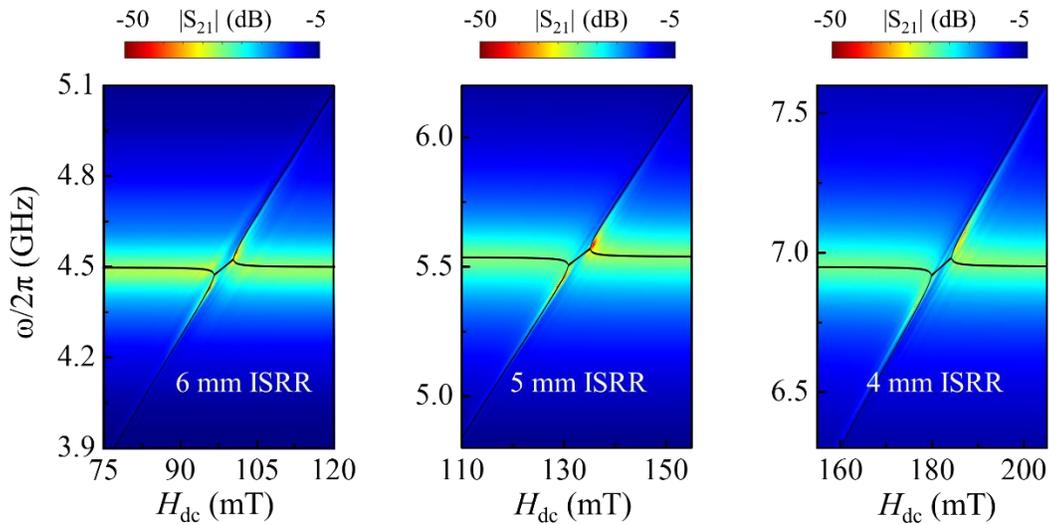

**Fig. S3.** Experimentally measured $|S_{21}|$ power on plane of $\omega/2\pi$ and $H_{\text{dc}}$ from hybrid samples consisting of YIG film and single ISRRs of (a) 6 mm, (b) 5 mm and (c) 4 mm size. The black solid lines correspond to the numerical calculations of Eq. (S8).

## S4. Analytical derivation for direct interaction of magnon with three photon modes

To understand the experimentally observed multiple coupling in the dispersion spectra due to the hybridization of magnons to multiple ISRRs' photon modes, we extended the analytical model of single-photon-magnon interaction to multi-photon-magnon interaction.

For the hybrid system shown in Fig. 1(a) of the main text, in the absence of coupling, when ac current $j$ is applied to the microstrip line, EMF voltage is generated in all the ISRRs. The angular resonance frequency and the damping of the n$^{th}$ ISRR are given by $\omega_n = 1/\sqrt{L_n C_n}$ and $\beta_n = R_n/L_n\omega_n$, respectively, where $L_n$, $C_n$ and $R_n$ are the inductance, capacitance and resistance of the n$^{th}$ ISRR, respectively.

Without coupling, the dynamic magnetization precession of YIG film is governed solely by the LLG equation. A more detailed description of how the LLG equation can be used to solve the magnetization precession and FMR is given in Ref [S2-S3]. Nonetheless, the solution for the magnetization precession using the LLG equation can be written as $m^+ = m_y + im_z = me^{-i\omega t}$. When YIG film is placed on the microstrip line, the magnetization precession induces, according to Faraday's induction law, an additional voltage to the n$^{th}$ ISRR: $(V_y)_n = (k_F)_n L_n (dm_z/dt)_n$ and $(V_z)_n = (k_F)_n L_n (dm_y/dt)_n$, where $(k_F)_n$ is the coupling parameter, due to Faraday's induction law. The total induced voltage $(V_n)_{p\leftarrow m} = (V_y)_n + i(V_z)_n = -(k_F)_n L_n \omega m^+$ can be added to the n$^{th}$ ISRR as an additional voltage source, which in turn changes the microwave current to the n$^{th}$ ISRR in the three concentric ISRRs to $(V_n)_{p\leftarrow m} = Z_n J_n^+$. Here, $Z_n$ is the impedance of the n$^{th}$ ISRR as given by $Z_n = -\frac{iL_n}{\omega}(\omega^2 - \omega_n^2 + 2i\beta_n\omega\omega_n)$. The equation for the coupling of the magnon mode to each photon mode of the three concentric ISRRs can be written as

$$i(k_F)_1\omega^2 m^+ + (\omega^2 - \omega_1^2 + 2i\beta_1\omega\omega_1)J_1^+ = 0$$

$$i(k_F)_2\omega^2 m^+ + (\omega^2 - \omega_2^2 + 2i\beta_2\omega\omega_2)J_2^+ = 0 \qquad (S10)$$

$$i(k_F)_3\omega^2 m^+ + (\omega^2 - \omega_3^2 + 2i\beta_3\omega\omega_3)J_3^+ = 0$$

The additional microwave current induced by the three concentric ISRRs (due to the effect of Lenz's law) also creates a strong microwave magnetic field around each of the ISRRs' split gap. Thus, in a three-concentric-ISRR/YIG hybrid system, the magnetizations in the YIG are influenced by the two fields, where one felicitates the coherence (Ampère's law) while the other one felicitates dissipative (Lenz's law) coupling rates between the photon and magnon modes. The competition between the two effects determines the type of anti-crossing, either normal or opposite, between the coupled modes. Finally, the LLG equation in the rotating frame is thus written as

$$(\omega - \omega_r + i\alpha\omega)m^+ - i\omega_m(k_A - k_L)_1 J_1^+ - i\omega_m(k_A - k_L)_2 J_2^+ - i\omega_m(k_A - k_L)_2 J_2^+ = 0, \qquad (S11)$$

where $J_n^+ = i(h^+)_n/(k_A - k_L)_n$ is the microwave current in the n$^{th}$ ISRR, resulting, via Ampere's and Lenz's laws, in the magnetic fields $(h_y)_n = (k_A - k_L)_n (J_z)_n$ and $(h_z)_n = (k_A - k_L)_n (J_y)_n$, where $k_A$ and $k_L$ are the coupling parameters due to Ampere's and Lenz's laws, respectively. To obtain the simultaneous solutions of Eqs. (S10) and (S11), the matrix form is rewritten as

$$\begin{bmatrix} \omega - \widetilde{\omega}_r & \omega_m K_1^2 & \omega_m K_2^2 & \omega_m K_3^2 \\ \omega_1 & \omega - \widetilde{\omega}_1 & 0 & 0 \\ \omega_2 & 0 & \omega - \widetilde{\omega}_2 & 0 \\ \omega_3 & 0 & 0 & \omega - \widetilde{\omega}_3 \end{bmatrix} \begin{bmatrix} m^+ \\ J_1^+ \\ J_2^+ \\ J_3^+ \end{bmatrix} = 0 \qquad (S12)$$

where $\widetilde{\omega}_n = \omega_n - i\beta_n\omega_n$, $\omega_n$ and $\beta_n$ are the resonance frequency and damping parameters of the ISRR modes (n=1, 2 and 3), respectively. Thus, the coupling constant between the magnon mode and the n$^{th}$ ISRR mode is given as $K_n^2 \cong (k_F)_n(k_A - k_L)_n = (k_c)_n^2 - (k_d)_n^2$, with $(k_F)_n(k_A)_n \cong$ $(k_c)_n^2$ and $(k_F)_n(k_L)_n \cong (k_d)_n^2$, where $(k_c)_n$ and $(k_d)_n$ are the coherent and dissipative coupling rates between the magnon and the n$^{th}$ photon mode. However, for the ISRR split gap along the microstrip line, the hybrid system always shows opposite anti-crossing due to dominant dissipative interaction ($k_d > k_c$) with a complex coupling constant. For a multi-mode PMC, the transmission spectrum is given as

$$S_{21} \propto \frac{\omega^2(\omega - \widetilde{\omega}_r)}{\begin{bmatrix} \omega - \widetilde{\omega}_r & \omega_m K_1^2 & \omega_m K_2^2 & \omega_m K_3^2 \\ \omega_1 & \omega - \widetilde{\omega}_1 & 0 & 0 \\ \omega_2 & 0 & \omega - \widetilde{\omega}_2 & 0 \\ \omega_3 & 0 & 0 & \omega - \widetilde{\omega}_3 \end{bmatrix}} \quad (S13)$$

## S5. Numerical calculation of different types of dispersion for magnon-mediated photon-photon coupling

Figure 3 in the main text explains the magnon-mediated photon-photon coupling, in which the individual photon modes are coupled directly to a single magnon mode, after which their hybridized modes interact with each other. The hybridized mode for each of the magnon-mediated photon-photon couplings can be written in the light of Eq. (S8) as

$$(E_\pm)_n = \left[\frac{(\widetilde{\omega}_r + \widetilde{\omega}_n)}{2} \pm \frac{1}{2}\sqrt{(\widetilde{\omega}_r - \widetilde{\omega}_n)^2 + 2\omega_m \omega_n (k_{MP}^2 + k_{PP}^2)_n}\right] \quad \text{(S14)}$$

where $\widetilde{\omega}_r = \omega_r - i(\alpha_{\text{in}} + \alpha_{\text{cp}})_n \omega_r$ and $\widetilde{\omega}_n = \omega_n - i\beta_{\text{in}_n} \omega_n$ are the complex frequencies of the magnon and individual photon modes, $k_{MP}$ is the direct PMC constant as given by $ik_{dc}$, and $k_{PP}$ is the magnon-mediated, indirect photon-photon coupling constant. The transmission spectrum for each coupled mode due to magnon-mediated photon-photon coupling is given by

$$|S_{21}|_n = \Gamma \frac{\omega^2 (\omega - \widetilde{\omega}_r)}{(\omega - \widetilde{\omega}_r)(\omega - \widetilde{\omega}_n) - \frac{1}{2}\omega_m \omega_n (k_{MP}^2 + k_{PP}^2)_n} \quad \text{(S15)}$$

We numerically calculated $|S_{21}|$ power on the $\omega/2\pi$ - $H_{\text{dc}}$ plane for the three different coupling regions of the three-concentric-ISRR/YIG hybrid system using Eq. (S15) along with the real and imaginary part of the eigenfrequencies using Eq. (S14), as shown in Fig. 4 of the main text. The calculation results agree well with the three different experimentally observed dispersions, the normal and opposite anti-crossing at 7.02 and 4.4 GHz, respectively, and the CIA at 5.58 GHz.

## S6. Analytical calculation of eigenfrequencies of magnon-mediated photon-photon coupling modes

The complex eigenvalues $E_\pm = \omega_\pm - i\Delta\omega_\pm$ for magnon mediated photon-photon coupling can be written from Eq. (S14) as

$$E_\pm = \frac{1}{2}\left[(\widetilde{\omega}_r + \widetilde{\omega}_p) \pm \sqrt{(\widetilde{\omega}_r - \widetilde{\omega}_p)^2 + 2\omega_m\omega_p K^2}\right], \tag{S16}$$

where $\widetilde{\omega}_r = \omega_r - i(\alpha_{\text{in}} + \alpha_{\text{cp}})\omega_r$ and $\widetilde{\omega}_p = \omega_p - i\beta\omega_p$ are the complex frequencies of the magnon and the photon modes and $K^2 = k_{MP}^2 + k_{PP}^2$ and $\alpha_{\text{eff}} = \alpha_{\text{in}} + \alpha_{\text{cp}}$. The energy difference between the two eigenmodes is given as

$$(E_+ - E_-)^2 = (\widetilde{\omega}_r - \widetilde{\omega}_r)^2 + 2\omega_m\omega_p K^2. \tag{S17}$$

At the common resonance frequency $\omega_r = \omega_p$, the real part of Eq. (S17) is transformed into

$$\left(\frac{\omega_+ - \omega_-}{\omega_p}\right)^2 = \left(\frac{\Delta\omega_+ - \Delta\omega_-}{\omega_p}\right)^2 - (\beta - \alpha_{\text{eff}})^2 + 2\frac{\omega_m}{\omega_p}K^2. \tag{S18}$$

The frequency gap $(\omega_+ - \omega_-)/2\pi$ at the coupling center is determined by the condition $(\Delta\omega_+ - \Delta\omega_-)/\omega_p \sim 0$, as

$$\Delta = \frac{\omega_+ - \omega_-}{2\pi} = \frac{1}{2\pi}\sqrt{2\omega_m\omega_p K^2 - \omega_p^2(\beta - \alpha_{\text{eff}})^2}. \tag{S19}$$

By separating the direct photon-magnon dissipative coupling and the magnon-mediated indirect photon-photon coupling, $K$ can be given as $K^2 = k_{MP}^2 + k_{PP}^2$. And, since $k_{MP}=ik$ for all of the ISRR/YIG hybrid samples, Eq. (S19) becomes

$$\Delta = \frac{1}{2\pi}\sqrt{2\omega_m\omega_p(-k^2 + k_{PP}^2) - \omega_p^2(\beta - \alpha_{\text{eff}})^2}. \tag{S20}$$

Now, given that the coupling constant $k_{pp}$ due to magnon-mediated indirect photon-photon coupling depends on the relative amplitude ($\sigma$) and phase difference ($\psi$) between the photon modes, $k_{pp}$ can be expressed as $k_{PP}^2 \simeq \sigma k^2 \cos\psi$. And, under the condition $\omega_m \simeq \omega_p$, Eq. (S20) becomes

$$\Delta = \frac{\omega_p}{2\pi}\sqrt{-2k^2(1-\sigma\cos\psi)-(\beta-\alpha_{\text{eff}})^2}. \tag{S21}$$

# Supplementary References